# Gate-tuned Aharonov-Bohm interference of surface states in a quasi-ballistic Dirac semimetal nanowire

Ben-Chuan Lin[1#], Shuo Wang[1#], Li-Xian Wang[1], Cai-Zhen Li[1], Jin-Guang Li[1], Dapeng Yu[1,2,3], Zhi-Min Liao[1,4]*

[1]State Key Laboratory for Mesoscopic Physics, School of Physics, Peking University, Beijing 100871, China

[2]Electron Microscopy Laboratory, School of Physics, Peking University, Beijing 100871, China

[3]Department of Physics, South University of Science and Technology of China, Shenzhen 518055, China

[4]Collaborative Innovation Center of Quantum Matter, Beijing 100871, China

#These authors contributed equally to this work.

*E-mail: liaozm@pku.edu.cn

**We report an observation of a topologically protected transport of surface carriers in a quasi-ballistic $Cd_3As_2$ nanowire. The nanowire is thin enough for the spin-textured surface carriers to form 1D subbands, demonstrating conductance oscillations with gate voltage even without magnetic field. The $\pi$ phase-shift of Aharonov-Bohm oscillations can periodically appear/disappear by tuning gate voltage continuously. Such a $\pi$ phase shift stemming from the Berry's phase demonstrates the topological nature of surface states. The topologically protected transport of the surface states is further revealed by four-terminal nonlocal measurements.**

Dirac semimetals [1], such as $Cd_3As_2$ or $Na_3Bi$ [2–10], show a linear electronic dispersion in three dimensions described by two copies of the Weyl equation. Applying a magnetic field can break the time reversal symmetry, and the Dirac semimetal is transformed into a Weyl semimetal with the two Weyl nodes separated in the momentum space [10,11]. Chiral charge pumping between the Weyl nodes with different chirality is predicted, which brings the Weyl fermions into the experimental realm. Recently, anomalous transport properties signaled by a pronounced negative magnetoresistance are observed as the evidence for the chiral anomaly effect [10,12]. Besides this, the surface dispersion-relation of a Weyl semimetal is topologically equivalent to a non-compact Riemann surface without equal-energy contour that encloses the projection of the Weyl point [13], leading to the emergence of surface Fermi arcs [14]. Lots of angle-resolved photoemission spectroscopy (ARPES) experiments [7,15–18] have confirmed the existence of Fermi arc, which connects different Weyl nodes in the projected 2D Brillouin zone. Moreover, the Fermi arcs have been demonstrated to be spin polarized [16,19] with a spin-polarization magnitude as large as 80% [19]. However, it is still hard to detect the properties of the topological surface states in transport experiments [20], mainly because the unavoidable conducting bulk states.

To reveal the transport properties of surface states, Aharonov-Bohm (AB) oscillations are particularly interesting phenomena to rule out the bulk contribution. Recent studies on AB oscillations in topological semimetals [21] have provided evidence of surface states conductions. The conductance oscillations [22] with flux quantum (h/e) periodicity were firstly demonstrated in metallic rings [23]. But notice that in metallic cylinders the h/e period AB oscillations cannot be observed due to the ensemble averaging of different slices of the metal cylinder [24,25]. The AB oscillations observed in topological nanoribbons are believed to originate from the quantum confinement and the circumferential interference of the surface states [26–31]. Moreover, the spin-helical nature of the topological surface states would introduce an additional Berry phase π as the carriers cycle along the perimeter [16,19,32–38]. Although the one-dimensional helical transport has been demonstrated in topological insulator nanowires through measuring the AB oscillations [31-33], the helical transport of the Dirac semimetal surface state is still a mysterious veil.

In this letter, we report on the topological transport of surface states of a single-crystalline $Cd_3As_2$ nanowire. The conductance is shown to oscillate with gate voltage even without magnetic field. When a magnetic field is applied, the AB oscillations as a function of magnetic field show a tunable phase shift by gate voltage. The AB and π phase-shift AB (π-AB) oscillations are further demonstrated by nonlocal measurements. The results are in consistent with theoretical explanation of the existence of 1D surface subbands and the spin-helical nature of the topological surface states.

The $Cd_3As_2$ nanowires were synthesized by chemical vapor deposition method [12]. The $Cd_3As_2$ nanowire is thin enough to have large surface-to-volume ratio and owns high crystal quality [12]. The scanning electron microscope (SEM) image of a typical device is shown in Fig. 1a. A typical device for the transport measurements comprised a thin nanowire with diameter of ~ 58 nm and a channel length ~80 nm between two voltage probes. Figure 1b shows the transfer curve of a nanowire device without magnetic field. Unlike preceding literatures [21,26–31], the transfer curve shows anomalous oscillations with back gate voltage $V_g$. Such anomalous oscillations may arise from the variation of the density of states (DOS) as tuning the Fermi level by gate voltage.

To further investigate the origin of such anomalous oscillations, we apply a magnetic field under a series of gate voltages. Two typical back gate voltages ($V_1$=-1V, $V_2$=-13V) are marked in Fig. 1b. The magnetic field modulations of the conductance are shown in Fig. 1c. The conductance oscillations $\Delta G$ versus magnetic flux at different gate voltages show a typical periodic behavior with the periodicity of $\Phi_0 = h/e$. $\Phi_0$ is the flux quantum and $\Phi_0 = \Delta B \cdot S$, where $\Delta B$ is the measured magnetic field periodicity ($\Delta B = 1.64T$ in this case) and $S$ is the cross-sectional area. From the magnetic field periodicity, we can deduce the cross-sectional area to be $2523 nm^2$, which is consistent with the nanowire diameter ~58 nm. In Fig. 1c, when $V_g$=-13 V, the conductance oscillations show conductance peaks at multiples of $\Phi_0$ and conductance valleys at half integer multiple of $\Phi_0$, a signature of the traditional AB oscillations [21,26–28]. However, when $V_g$=-1 V, opposite conductance oscillations are observed, that is, the peaks are located at half integer multiple of $\Phi_0$ and the valleys at integer multiple of $\Phi_0$, showing a π-AB oscillations [29–31]. These quantum oscillations were even observed under high magnetic field up to 10 T (Fig. 1c).

To further clearly present the conductance oscillations, we plot the mapping of $\Delta G$ versus gate voltage and magnetic field in Fig. 1d. Clearly there are two kinds of phase modulations on the interference. One is tuned by gate voltage, and the other is influenced by the magnetic field. At a fixed gate voltage, if the conductance reaches the minimum at zero magnetic field, the conductance will be the maximum at half integer multiple of $\Phi_0$; if the conductance is maximum at zero magnetic field, the conductance will be the maximum at integer multiple of $\Phi_0$. The phase of the AB interference is strongly dependent on gate voltage.

The π phase shift AB oscillations have been theoretically predicted [35,37,38] and experimentally investigated in topological insulators (TI) nanostructures [28–31]. In the presence of an axial magnetic field, the band structure of surface carriers in a TI nanostructure can be described by [30,31,35,37,38]

$$E = \pm h v_F \sqrt{\frac{k^2}{4\pi^2} + \left(\frac{m+\frac{1}{2}-\frac{\Phi}{\Phi_0}}{C}\right)^2}, \quad (1)$$

where $h$ is the Planck constant, $v_F$ is the Fermi velocity, the momentum vector $k$ is along the nanowire/nanoribbon axis, $m$ is the angular momentum and $C$ is the circumference of the nanostructure. The 1/2 term comes from the Berry phase π [16,19,32–34]. When the AB phase term (namely $\Phi/\Phi_0$) equals half integer, a gapless mode can be realized. Otherwise there would be a gap near the Dirac point. This model successfully explain the π-AB oscillations observed in the TI nanostructures [29–31].

In this work, considering we have observed both the AB oscillations and π-AB oscillations, we propose that such a model in TIs can also be modified to apply to the topological semimetals. Things happen to be some different in topological semimetals, because the Dirac point in Dirac semimetal is two-fold degenerate of Weyl nodes. The corresponding Weyl equation can be simply described [39] as $E = \pm \hbar v_F \cdot k$, where $\pm$ denotes different chirality. For example, if the chirality is +1, the original energy dispersion relation $E = +\hbar v_F \cdot k$ and the surface band splitting modification would become

$$E = \begin{cases} +h v_F \sqrt{\frac{k_\parallel^2}{4\pi^2} + \left(\frac{m+\frac{1}{2}-\frac{\Phi}{\Phi_0}}{C}\right)^2}, & for\ k_\parallel \geq 0 \\ -h v_F \sqrt{\frac{k_\parallel^2}{4\pi^2} + \left(\frac{m+\frac{1}{2}-\frac{\Phi}{\Phi_0}}{C}\right)^2}, & for\ k_\parallel \leq 0 \end{cases} \quad for\ chirality +1, \quad (2)$$

while when the chirality is -1, the energy dispersion has a similar form with a sign change. This physics picture is depicted in Fig. 2. At zero magnetic field, that is $\Phi =0$, the original linear energy dispersion becomes gapped with a series of sub-bands, as shown in Fig. 2a. The red and blue lines represent the chirality to be +1 and -1, respectively. According to the surface band splitting, there should emerge a periodic oscillation when the Fermi level crosses the sub-bands continuously. This is what happens in our $Cd_3As_2$ nanowires, as shown in Fig. 1b.

When a magnetic field is applied, the corresponding AB oscillation term $\Phi/\Phi_0$ should be considered. The surface energy band diagrams at the magnetic flux $\Phi = m\Phi_0$ and $\Phi = (m+\frac{1}{2})\Phi_0$ are depicted in Fig. 2b, where the letter $L$ and $R$ denote the chirality of Weyl nodes to be +1 or −1. Apparently, when the magnetic flux is half integer of $\Phi_0$, the linear energy band with specified chirality emerges. The quantum transport can be co-modulated by both gate voltage and magnetic field. From the energy band diagram, there will be a conductance minimum at $\Phi = m\Phi_0$ and a conductance maximum at $\Phi = (m+\frac{1}{2})\Phi_0$ when the Fermi level is near the Wely node or at the positions indicated by the black horizontal lines in Fig. 2b. In contrast, when $E_F$ is located at the positions indicated by the pink dashed lines in Fig. 2b, there will be a conductance peak at $\Phi = m\Phi_0$ and a conductance valley at $\Phi = (m+\frac{1}{2})\Phi_0$, because the parabolic bands are two-fold degenerate while the linear bands are nondegenerate. According to this physical picture, the conductance oscillations co-modulated by magnetic flux and gate voltage in Fig. 1d are well understood.

According to equation 2, the gap energy of the surface sub-bands is $\Delta \sim \hbar v_F\sqrt{\pi/S}$ [29–31,35]. By considering a typical value of Fermi velocity $v_F = 1 \sim 5 \times 10^5 m/s$ for $Cd_3As_2$, the gap energy is estimated to be $2.32 \sim 11.63\ mev$, corresponding to $27 \sim 153\ K$. Thus we can expect such periodic oscillations at high temperatures, although the experimental temperature is usually below 1 $K$ in previous literatures [29–31]. Figures 3a,b present the mapping plots of $\Delta G$ versus magnetic field and temperature $T$, with fixed $V_g$ of -1 V and -5 V, respectively. The two types of oscillations are both very clear at high temperature of 22 $K$, which confirms the large energy gap between surface sub-bands. The temperature dependence of $L_\Phi$ can be obtained from the fast-Fourier-transform (FFT) of the oscillations by $A_{FFT} \sim e^{-\alpha/L_\Phi(T)}$, where $A_{FFT}$ is the amplitude of FFT, and $\alpha$ is a constant [40]. The linear behaviors ($\ln A_{FFT} \sim -\alpha T$) presented in Figs. 3c,d indicate the $L_\Phi \sim T^{-1}$ dependence. The $T^{-1}$ dependence of $L_\phi$ suggests the quasi-ballistic transport [28], and the decoherence is limited by a weak coupling to the fluctuating environment [41].

To further reveal the topological properties of the surface states, we have performed the four-terminal nonlocal measurements. As schematized in Fig. 4a, the current is applied along two terminals and the nonlocal voltage on the adjacent two probes is measured. The gate voltage dependent conductance oscillations under zero magnetic field are clearly observed from the nonlocal measurements, as shown in Fig. 4b. It can be seen that the nonlocal signals show consistent oscillations with the local signals, demonstrating the topological robustness of the transport without phase losing. Since $B = 0$ and the chiral anomaly $E \cdot B$ term is zero, it rules out the possible chiral anomaly effect induced nonlocal transport from the bulk channel [42].

With a magnetic field applied, the conductance oscillations measured from the local two-probe ($\Delta G_L$) and the non-local configurations ($\Delta G_{NL}$) are presented in Fig. 4c and Fig. 4d, respectively. The nonlocal oscillations are still highly consistent with the local oscillations, demonstrating nearly no phase losing during the nonlocal transport as the surface states are topologically protected.

In Fig. 4c, small peaks near zero magnetic field are clearly observed in the local region while these peaks disappear in the nonlocal regime as shown in Fig. 4d. Considering the large spin-orbital coupling strength in Cd3As2 sample, such conductance peaks may come from the WAL effects [43–46] which is dominated by bulk states in this case. In the nonlocal regime, with less bulk states contribution, such peaks are smeared out by the π-AB effect induced by surface states.

In conclusion, we have demonstrated both local and nonlocal transport phenomena dominated by topological surface states of individual Dirac semimetal nanowires. The thin nanowires have large surface-to-volume ratio and quantum confinement induced 1D surface sub-bands. Anomalous quantum oscillations are clearly observed by tuning the Fermi level *via* gate voltage across each surface sub-bands without magnetic field. The surface sub-bands are further tunable to re-present a linear helical mode at $(m+\frac{1}{2})\frac{h}{e}$ magnetic flux, giving rise to conductance peaks as the Fermi level is near the Dirac point. Such helical transport of the surface states also shows quasi-ballistic behavior and is further confirmed by the nonlocal transport measurements.

*Acknowledgement.* This work was supported by National Key Research and Development Program of China (Nos. 2016YFA0300802, 2013CB934600, 2013CB932602) and NSFC (Nos. 11274014, 11234001).

# REFERENCES

[1] T. O. Wehling, A. M. Black-Schaffer, and A. V. Balatsky, Adv. Phys. **63**, 1 (2014).

[2] J. L. Mañes, Phys. Rev. B **85**, 155118 (2012).

[3] S. M. Young, S. Zaheer, J. C. Y. Teo, C. L. Kane, E. J. Mele, and A. M. Rappe, Phys. Rev. Lett. **108**, 140405 (2012).

[4] Z. Wang, Y. Sun, X.-Q. Chen, C. Franchini, G. Xu, H. Weng, X. Dai, and Z. Fang, Phys. Rev. B **85**, (2012).

[5] Z. Wang, H. Weng, Q. Wu, X. Dai, and Z. Fang, Phys. Rev. B **88**, 125427 (2013).

[6] T. Liang, Q. Gibson, M. N. Ali, M. Liu, R. J. Cava, and N. P. Ong, Nat. Mater. **14**, 280 (2014).

[7] S.-Y. Xu, C. Liu, S. K. Kushwaha, R. Sankar, J. W. Krizan, Ilya Belopolski, Madhab Neupane, Guang Bian, Nasser Alidoust, Tay-Rong Chang, Horng-Tay Jeng, Cheng-Yi Huang, Wei-Feng Tsai, Hsin Lin, Pavel P. Shibayev, Fang-Cheng Chou, Robert J. Cava, and M. Zahid Hasan, Science **347**, 294 (2015).

[8] Z. K. Liu, J. Jiang, B. Zhou, Z. J. Wang, Y. Zhang, H. M. Weng, D. Prabhakaran, S.-K. Mo, H. Peng, P. Dudin, T. Kim, M. Hoesch, Z. Fang, X. Dai, Z. X. Shen, D. L. Feng, Z. Hussain, and Y. L. Chen, Nat. Mater. **13**, 677 (2014).

[9] Z. K. Liu, B. Zhou, Y. Zhang, Z. J. Wang, H. M. Weng, D. Prabhakaran, S.-K. Mo, Z. X. Shen, Z. Fang, X. Dai, Z. Hussain, and Y. L. Chen, Science **343**, 864 (2014).

[10] X. Jun, K. K. Satya, L. Tian, W. K. Jason, H. Max, W. Wudi, C. R. J., and O. N. P., Science **350**, 413 (2015).

[11] von H. Weyl, Z Phys. (1929).

[12] C.-Z. Li, L.-X. Wang, H. Liu, J. Wang, Z.-M. Liao, and D.-P. Yu, Nat. Commun. **6**, 10137 (2015).

[13] C. Fang, L. Lu, J. Liu, and L. Fu, Nat. Phys. (2016).

[14] X. Wan, A. M. Turner, A. Vishwanath, and S. Y. Savrasov, Phys. Rev. B **83**, 205101 (2011).

[15] B. Q. Lv, H. M. Weng, B. B. Fu, X. P. Wang, H. Miao, J. Ma, P. Richard, X. C. Huang, L. X. Zhao, G. F. Chen, Z. Fang, X. Dai, T. Qian, and H. Ding, Phys. Rev. X **5**, 031013 (2015).

[16] B. Q. Lv, S. Muff, T. Qian, Z. D. Song, S. M. Nie, N. Xu, P. Richard, C. E. Matt, N. C. Plumb, L. X. Zhao, G. F. Chen, Z. Fang, X. Dai, J. H. Dil, J. Mesot, M. Shi, H. M. Weng, and H. Ding, Phys. Rev. Lett. **115**, 217601 (2015).

[17] S.-Y. Xu, B. Ilya, Nasser Alidoust, Madhab Neupane, Guang Bian, Chenglong Zhang, Raman Sankar, Guoqing Chang, Zhujun Yuan, Chi-Cheng Lee, Shin-Ming Huang, Hao Zheng, Jie Ma, Daniel S. Sanchez, BaoKai Wang, Arun Bansil, Fangcheng Chou, Pavel P. Shibayev, Hsin Lin, Shuang Jia, and M. Zahid Hasan, Science **349**, 613 (2015).

[18] S.-Y. Xu, N. Alidoust, I. Belopolski, Z. Yuan, G. Bian, T.-R. Chang, H. Zheng, V. N. Strocov, D. S. Sanchez, G. Chang, C. Zhang, D. Mou, Y. Wu, L. Huang, C.-C.

Lee, S.-M. Huang, B. Wang, A. Bansil, H.-T. Jeng, T. Neupert, A. Kaminski, H. Lin, S. Jia, and M. Zahid Hasan, Nat. Phys. **11**, 748 (2015).

[19] S.-Y. Xu, I. Belopolski, D. S. Sanchez, M. Neupane, G. Chang, K. Yaji, Z. Yuan, C. Zhang, K. Kuroda, G. Bian, C. Guo, H. Lu, T.-R. Chang, N. Alidoust, H. Zheng, C.-C. Lee, S.-M. Huang, C.-H. Hsu, H.-T. Jeng, A. Bansil, T. Neupert, F. Komori, T. Kondo, S. Shin, H. Lin, S. Jia, and M. Z. Hasan, Phys. Rev. Lett. **116**, (2016).

[20] P. J. W. Moll, N. L. Nair, T. Helm, A. C. Potter, I. Kimchi, A. Vishwanath, and J. G. Analytis, Nature **535**, 266 (2016).

[21] L.-X. Wang, C.-Z. Li, D.-P. Yu, and Z.-M. Liao, Nat. Commun. **7**, 10769 (2016).

[22] Y. Aharonov and D. Bohm, Phys. Rev. **115**, 485 (1959).

[23] S. Washburn and R. A. Webb, Adv. Phys. **35**, 375 (1986).

[24] C. P. Umbach, C. Van Haesendonck, R. B. Laibowitz, S. Washburn, and R. A. Webb, Phys. Rev. Lett. **56**, 386 (1986).

[25] M. Murat, Y. Gefen, and Y. Imry, Phys. Rev. B **34**, 659 (1986).

[26] H. Peng, K. Lai, D. Kong, S. Meister, Y. Chen, X.-L. Qi, S.-C. Zhang, Z.-X. Shen, and Y. Cui, Nat. Mater. **9**, 225 (2009).

[27] F. Xiu, L. He, Y. Wang, L. Cheng, L.-T. Chang, M. Lang, G. Huang, X. Kou, Y. Zhou, X. Jiang, Z. Chen, J. Zou, A. Shailos, and K. L. Wang, Nat. Nanotechnol. **6**, 216 (2011).

[28] J. Dufouleur, L. Veyrat, A. Teichgräber, S. Neuhaus, C. Nowka, S. Hampel, J. Cayssol, J. Schumann, B. Eichler, O. G. Schmidt, B. Büchner, and R. Giraud, Phys. Rev. Lett. **110**, 186806 (2013).

[29] S. S. Hong, Y. Zhang, J. J. Cha, X.-L. Qi, and Y. Cui, Nano Lett. **14**, 2815 (2014).

[30] S. Cho, B. Dellabetta, R. Zhong, J. Schneeloch, T. Liu, G. Gu, M. J. Gilbert, and N. Mason, Nat. Commun. **6**, 7634 (2015).

[31] L. A. Jauregui, M. T. Pettes, L. P. Rokhinson, L. Shi, and Y. P. Chen, Nat. Nanotechnol. **11**, 345 (2016).

[32] D. Xiao, M.-C. Chang, and Q. Niu, Rev. Mod. Phys. **82**, 1959 (2010).

[33] M. Z. Hasan and C. L. Kane, Rev. Mod. Phys. **82**, 3045 (2010).

[34] X.-L. Qi and S.-C. Zhang, Rev. Mod. Phys. **83**, 1057 (2011).

[35] G. Rosenberg, H.-M. Guo, and M. Franz, Phys. Rev. B **82**, (2010).

[36] Y. Ran, A. Vishwanath, and D.-H. Lee, Phys. Rev. Lett. **101**, (2008).

[37] J. H. Bardarson, P. W. Brouwer, and J. E. Moore, Phys. Rev. Lett. **105**, (2010).

[38] Y. Zhang and A. Vishwanath, Phys. Rev. Lett. **105**, (2010).

[39] Y. Baum, E. Berg, S. A. Parameswaran, and A. Stern, Phys. Rev. X **5**, 041046 (2015).

[40] A. E. Hansen, A. Kristensen, S. Pedersen, C. B. SHrensen, and P. E. Lindelof, Phys. E 770 (2002).

[41] G. Seelig and M. Büttiker, Phys. Rev. B **64**, 245313 (2001).

[42] S. A. Parameswaran, T. Grover, D. A. Abanin, D. A. Pesin, and A. Vishwanath, Phys. Rev. X **4**, 031035 (2014).

[43] H.-J. Kim, K.-S. Kim, J.-F. Wang, M. Sasaki, N. Satoh, A. Ohnishi, M. Kitaura,

M. Yang, and L. Li, Phys. Rev. Lett. **111**, 246603 (2013).

[44] C.-L. Zhang, S.-Y. Xu, I. Belopolski, Z. Yuan, Z. Lin, B. Tong, G. Bian, N. Alidoust, C.-C. Lee, S.-M. Huang, T.-R. Chang, G. Chang, C.-H. Hsu, H.-T. Jeng, M. Neupane, D. S. Sanchez, H. Zheng, J. Wang, H. Lin, C. Zhang, H.-Z. Lu, S.-Q. Shen, T. Neupert, M. Zahid Hasan, and S. Jia, Nat. Commun. **7**, 10735 (2016).

[45] Y. Liu, C. Zhang, X. Yuan, T. Lei, C. Wang, D. Di Sante, A. Narayan, L. He, S. Picozzi, S. Sanvito, R. Che, and F. Xiu, NPG Asia Mater. **7**, e221 (2015).

[46] B. Zhao, P. Cheng, H. Pan, S. Zhang, B. Wang, G. Wang, F. Xiu, and F. Song, Sci. Rep. **6**, (2016).

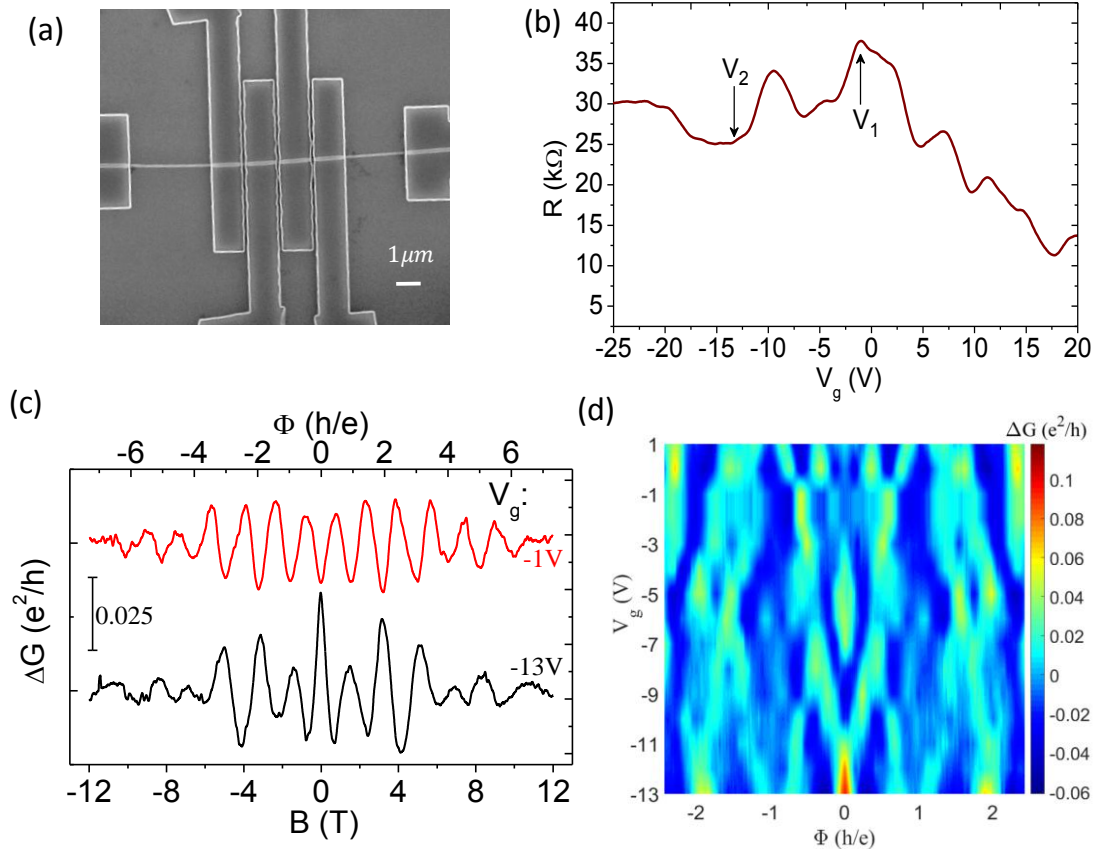

**Figure 1. (a)** SEM image of a typical $Cd_3As_2$ nanowire device. **(b)** The resistance versus $V_g$ measured from a four-terminal configuration. It shows evident oscillations tuned by $V_g$ as Fermi level $E_F$. Two gate voltages are masked with $V_1 = -1V$, and $V_2 = -13V$. **(c)** The conductance oscillations versus magnetic field at different gate voltages as marked in (b). **(d)** The mapping of conductance oscillation $\Delta G$ versus magnetic flux and gate voltage.

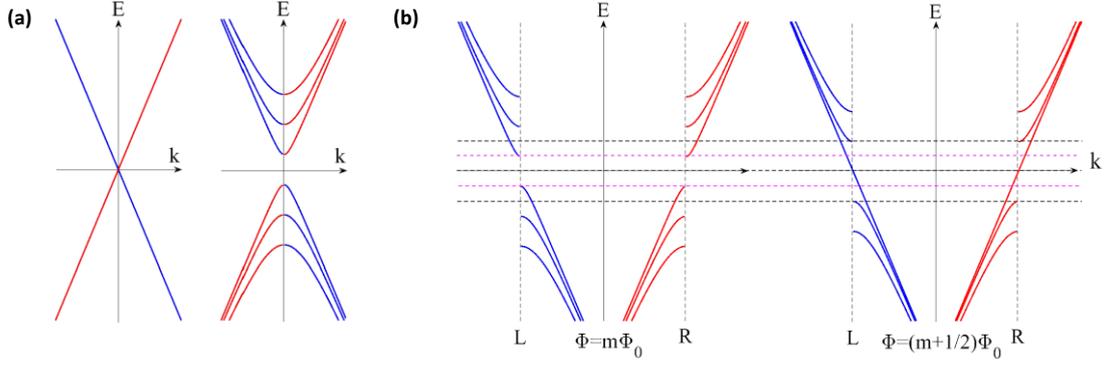

**Figure 2. Schematics of the surface energy band with/without magnetic field.** (**a**) No magnetic field is applied. The left sketch shows the original energy momentum dispersion relationship. The right panel shows the surface band splitting due to quantum confinement. The red and blue lines correspond to different chirality. (**b**) The surface energy band diagrams at the magnetic flux $\Phi = m\Phi_0$ and $\Phi = (m+\frac{1}{2})\Phi_0$, respectively. The Dirac point is separated into two Weyl nodes by the magnetic field. The letters *L, R* denote chirality to be −1 or +1. Linear Weyl modes appear at $\Phi = (m+\frac{1}{2})\Phi_0$. Note that only the linear Weyl mode is nondegenerate while other bands are two-fold degenerate of angular momentum. The black dashed lines indicate the Fermi level positions where the $\pi$ AB effect appears; and the pink dashed lines indicate the Fermi level positions for the 0 AB effect.

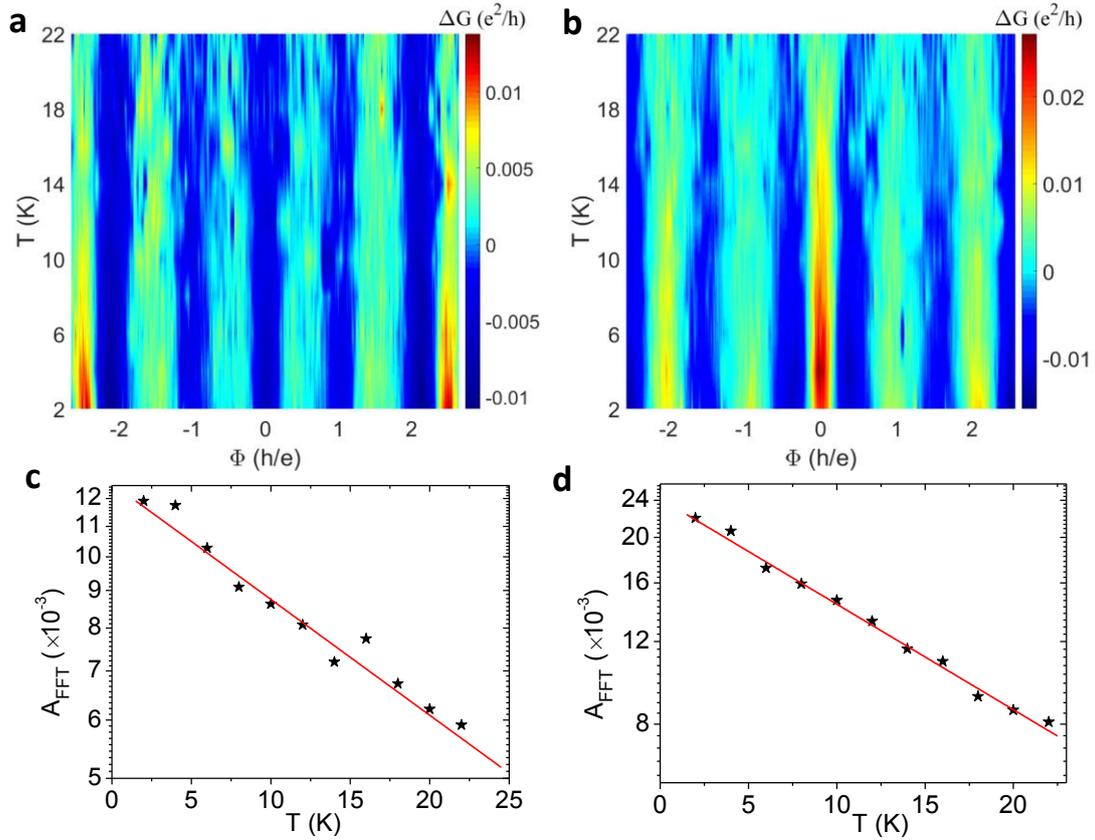

**Figure 3. Temperature and magnetic field dependence of conductance oscillations.** (**a**) The mapping of conductance oscillation $\Delta G$ versus magnetic field and temperature at the charge neutrality point, namely $V_g = -1V$. (**b**) The mapping of $\Delta G$ at $V_g = -5$V. The quantum oscillations are rather robust against temperature, indicating the topological nature. (**c, d**) The semilog plot of the amplitude $A_{FFT}$ versus temperature at $V_g = -1 \, and -5V$, respectively. The red line is the linear fitting.

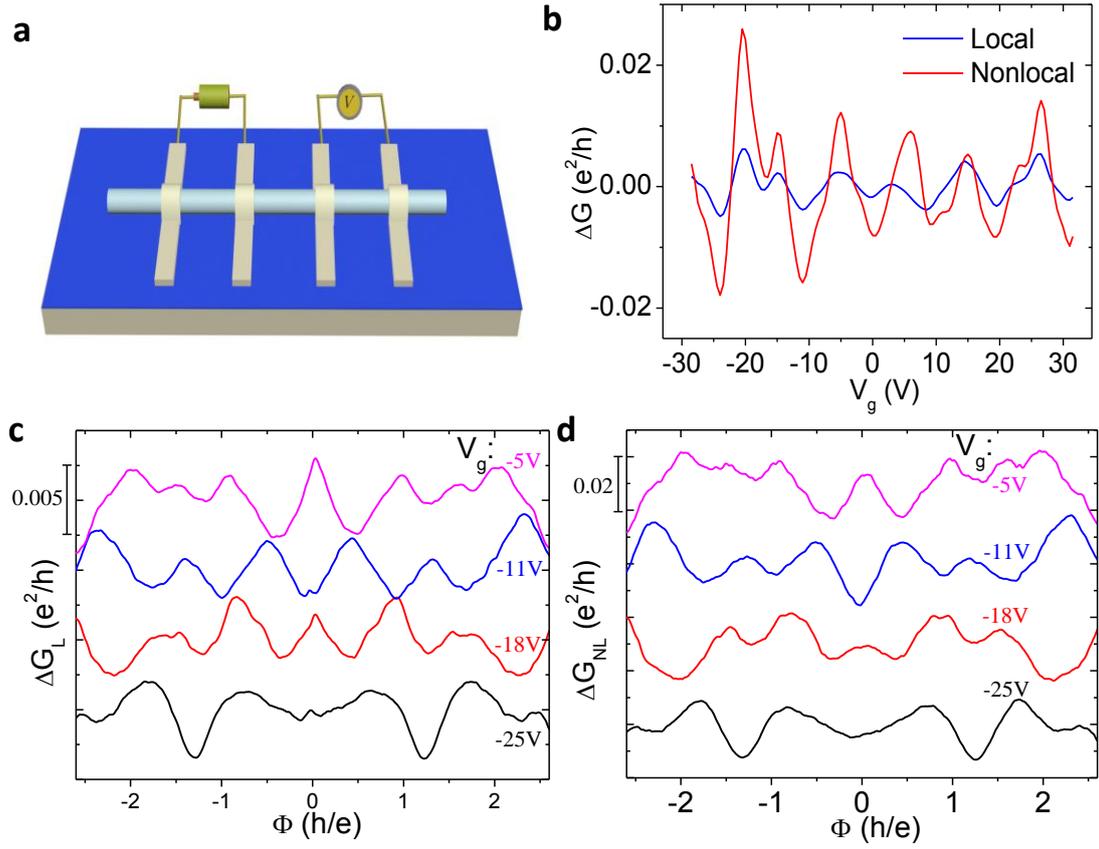

**Figure 4. Local and nonlocal signals of quantum oscillations. (a)** A sketch of the four-terminal nonlocal configuration. **(b)** Conductance oscillations tuned by gate voltage under zero magnetic field with both local and nonlocal configurations. **(c)** The local conductance oscillation Δ$G$ versus magnetic flux at different gate voltages. **(d)** The nonlocal conductance oscillation Δ$G$ versus magnetic flux at different gate voltages.